\pdfoutput=1
\documentclass[12pt]{article}
\usepackage[utf8]{inputenc}
\usepackage[margin=1in]{geometry}

\usepackage[auth-sc, affil-it]{authblk}
\usepackage{graphicx}
\usepackage{array}
\usepackage{longtable}   %
\usepackage{afterpage}   %

\makeatletter
\renewcommand{\fps@table}{tp}
\renewcommand{\fps@figure}{tp}
\makeatother

 \usepackage[doublespacing]{setspace}
\usepackage{fullpage}

\usepackage{natbib}
\usepackage{har2nat}
\usepackage{mathptmx}
\usepackage{footmisc}

 \usepackage{caption}
 \usepackage{subcaption}
 \usepackage{xcolor}
\definecolor{MyDarkBlue}{rgb}{0,0.08,0.45}
\usepackage{CJKutf8}   %

\usepackage{url}
\usepackage[colorlinks=true,allcolors=MyDarkBlue]{hyperref}
\usepackage{xurl}   %

\AtBeginDocument{%
  \renewcommand\harvardurl[1]{\textbf{URL:} \begingroup\urlstyle{same}\itshape\url{#1}\endgroup}%
}

 \graphicspath{{figs/}}
 \usepackage{bibunits}
\defaultbibliographystyle{apsr}
\defaultbibliography{refs.bib}

\newcolumntype{P}[1]{>{\raggedright\arraybackslash}p{#1}}

\makeatletter
\newif\if@gateon \@gateontrue
\newcommand{\gateoff}{\@gateonfalse}
\AtEndDocument{%
  \if@gateon
    \ifx\@multiplelabels\relax\else
      \PackageError{buildgate}{%
        Two \string\label s share a name%
      }{Search the log for 'multiply defined'. Any \string\ref\space to that^^J%
        name resolves to whichever LaTeX read last. Rename one, or^^J%
        \string\gateoff\space to bypass.}%
    \fi
  \fi
}
\makeatother

\hypersetup{
  pdftitle={Technology-Supported Moderation in Online Deliberation},
  pdfauthor={Lodewijk L. Gelauff, Ashish Goel, Sukolsak Sakshuwong}
}

\title{Technology-Supported Moderation in Online Deliberation}
\author{
    {\protect\href{https://orcid.org/0000-0003-1936-638X}{\protect\includegraphics[scale=0.06]{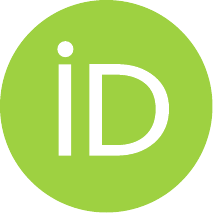}\protect\hspace{1mm}Lodewijk L. Gelauff} \thanks{
        Generation Lab, and Center on Democracy, Development and the Rule of Law,
        Stanford University. Corresponding author: \texttt{lodewijk@stanford.edu}.}},
    {Ashish Goel \thanks{
	   Management Science and Engineering, 	Stanford University. 	\texttt{ashishg@stanford.edu}}}, and
    {Sukolsak Sakshuwong \thanks{
        Google. Work was performed while at Stanford University. \texttt{sukolsak@google.com}
    }}
}

\date{October 2026\thanks{This version is the manuscript as submitted in February 2026, before editorial review, with stylistic and linguistic corrections and data corrections and updates made independently of the editorial process.}}

\begin{document}

\maketitle

\begin{quote}\small\singlespacing
This is a preprint of the following chapter: Lodewijk L. Gelauff, Ashish Goel and Sukolsak Sakshuwong, Technology-Supported Moderation in Online Deliberation, to appear in Democratizing Tech Governance: Empowering the Public's Voice Through Deliberative Democracy (preliminary title), edited by Alice Siu, forthcoming%
. It is the version of the author's manuscript prior to acceptance for publication and has not undergone editorial and/or peer review on behalf of the Publisher (where applicable).
\end{quote}

\begin{abstract}
\singlespacing
Mini-publics are an increasingly popular family of democratic innovation to broaden citizen engagement. However, convening a representative sample in person is resource-intensive, and requires significant organizational efforts to coordinate travel, venue logistics, and facilitator training.
This chapter outlines the design and development of the Stanford Online Deliberation Platform, initially tailored for Deliberative Polling, building upon insights from in-person moderation in that setting, but supporting a broader range of methods and approaches in the deliberative space.

The platform supports structured conversations for small groups (ideally 8-15 people) with automated moderation functionality, enabling the groups to assume ownership of the deliberative conversation rather than relying on a moderator persona. We describe the platform's architecture and key design decisions, including the deliberation flow, room assignment, speaker management, speaking nudges, agenda management, outcome development, and real-time monitoring.

Between its inception in 2018 and 2025, the platform has facilitated over 30,000 unique participants engaging in deliberations; we catalog 54 deliberations with at least 100 participants up to December 2025, spanning a diverse range of languages, countries, and topics. Finally, we establish a standardized vocabulary for future analyses by defining the terminology employed in the platform's data and documenting significant software revisions over time.
\end{abstract}

\begin{bibunit}

\section{Introduction}
\label{sec:introduction}
Many democratic innovation initiatives have leveraged deliberation through mini-publics over the past decades in an effort to broaden citizen engagement in public policy decision-making. Whether using deliberative democracy to evaluate a proposed constitutional reform or to debate trade-offs within AI governance, a remarkable phenomenon seems to occur when representatives from across society convene to thoughtfully discuss a complex topic: participants in the mini-public engage in ways that might defy the expectations shaped by today's polarized media landscape. After having hosted dozens of these deliberations on the Stanford Online Deliberation Platform between 2018 and 2025, this chapter provides a detailed description of the platform's structure, key design choices and an overview of the deliberations that we hosted.

In Deliberative Polling, James Fishkin has demonstrated the capacity of mini-publics to reveal a hypothetical `informed public opinion' on complex issues: what would people think about a topic, if they were able to consider its facets under \textit{good conditions}. Participants in these polls often shift their policy preferences significantly, increase their knowledge on the issues and show greater empathy to those they disagree with \cite{fishkin_democracy_2019, fishkin_is_2021, fishkin_can_2024}.

Fishkin's Center for Deliberative Democracy (later the Deliberative Democracy Lab) organized over 50 in-person offline deliberations between 1990 and 2017, including a poll in Texas on energy policy, a US National Issues Convention and a consultation on a constitutional change in Mongolia.
These exercises in deliberative democracy required convening a representative sample in a single location with small-group discussions facilitated in a neutral, observational approach.
Besides participant compensation, organizers faced substantial costs and organizational effort for arranging travel, venue and facilitator training.
These required resources alone would motivate efforts to consider transitioning to deliberations online. It should therefore not be surprising that our platform was not the first attempt to make this possible \cite{coleman_under_2012}. A pathway to organize deliberations reliably online with similar outcomes would not only bring down the costs of these exercises, but also expand accessibility to a wider range of applications.

This chapter details the design and evolution of an online platform purpose-built for Deliberative Polling, focusing on its technical infrastructure and methodological framework. By systematically documenting its architecture, decision-making processes, and iterative improvements, we aim to establish a shared vocabulary for analyzing the platform's role in enabling deliberative experiments. The chapter proceeds as follows: First, we briefly situate the platform within scholarly debates about digital deliberation, and other publications that discuss the platform in some way. We will list the deliberations that have taken place on the platform, and outline their characteristics, and place them in time. Next, we outline its core functionalities. Finally we will share a developmental timeline in the Appendix to allow the reader to deduce the state of the platform for each deliberation.

Deliberative, structured, small-group conversations are the broader use-case that this platform was designed for. To illustrate the richness of this spectrum, consider the following design considerations:
\begin{itemize}
    \item Group dynamics: The size of the groups where the `deliberative action' takes place, is perhaps the most defining choice for an organizer. Interaction between participants can be structured as plenary discussions where each participant receives very limited speaking time, smaller groups (8-15 people) similar to break-out rooms, pairwise exchanges, or even individual exchanges with an AI interface.
    \item Temporal dynamics: Are the participants communicating synchronously or asynchronously?
    \item Communication medium: Are participants communicating primarily through written contributions or primarily through spoken contributions?
    \item Stated objective: What is the purpose of the exercise? Examples include consensus building, proposal generation, and open-ended perspective exchange.
\end{itemize}

Not only do these choices have an enormous impact on how such an exchange would be facilitated, but they would also have very meaningful consequences for the logistics that the organizer is confronted with, and how participants can be recruited. Whether participants have to be all available at the same time (synchronous) or are asked to repeatedly check a communication platform (asynchronous) has for example large implications on recruitment, as the availability rapidly becomes a determining factor whether people are available to participate in the first place.

To make this section somewhat manageable, we will limit ourselves to the choices that would be made for a deliberation in the context of Deliberative Polling: small-group (8-15 people), synchronous, spoken deliberation where participants are exchanging viewpoints but not asked to achieve consensus or convince each other.

\subsection{Why online?}
\label{sec:why-online}
Organizing a synchronous deliberation with a mini-public in an offline setting often requires substantial investments of time and resources by both organizers and participants.
Shifting a deliberation to an online setting can reduce required resources, while expanding opportunities for data collection, analysis, scale and allow a broader range of topics and contexts where deliberative mini-publics could be considered.

In this section, we discuss factors that organizers could consider when deciding between online and in-person deliberation.

\subsubsection{Logistics and Cost}
For organizers, adopting an online platform may simplify logistical complexity. Eliminating the need to book travel, venue, accommodation and catering substantially reduces logistical burden and associated costs.
The shift also removes the need to dedicate valuable staff time to coordinate these logistical tasks, reducing or removing a bottleneck in the organizing timeline for the organizing team. This frees up their time for other tasks, but also opens up additional opportunities to be pursued.

For potential participants, online deliberation may reduce the time and effort they would have to commit in order to be able to accept the invitation. Time investment is limited to the scheduled sessions, surveys and preparation time, allowing the participants to manage the use of break time for personal needs and obligations more flexibly. The lower time commitment due to the absence of travel reduces the need for extensive personal arrangements.
For some potential participants with family, mobility or health constraints, an online option may remove concerns to arrange for their personal needs in order to participate.

These efficiencies may enhance organizers' ability to recruit certain demographic intersections as geographic and scheduling constraints are less impactful.

\subsubsection{Flexibility}
The online format extends the range of possible agenda configurations and designs within the same total deliberation time. For example, some online deliberations structured their sessions across multiple weekday evenings or weekend days over several weeks. This allowed the organizers to tailor to the needs and availability of the participants in specific regions or demographics that were otherwise harder to recruit.

Online deliberation reduces the fixed costs associated with the in-person logistics, making it feasible to organize deliberations in order to address narrower policy questions. As a traditional in-person deliberation in a country like the United States may have required the organizers to bundle multiple topics into a weekend-long event to justify logistical investments from organizers and participants alike, online formats allow organizers to design focused 2-hour sessions with small-group discussions and an expert panel. Recruitment costs and overall coordination remain as fixed expenses, while participant compensation and agenda design become more modular and scalable.

Conversely, using an online platform also allows the organizer to consider more expansive and complex agenda designs, spanning multiple weekends or small-group discussions with recruitment through a panel design -- a possibility yet to be fully explored in practice.

\subsubsection{Taking the mini out of a mini-public}
\citet{ackerman_deliberation_2004} proposed a ``Deliberation Day''  to strengthen democratic discourse by enabling citizens to deliberate on key issues immediately before an election. Building on this premise we could explore the very act of deliberating as a civic exercise training the democratic muscle.
Many designs are possible where not (only) a mini-public deliberates, but the practice expands beyond a representative sample to engage broader populations in structured conversation. This kind of scaling could, under the right conditions, allow larger groups of participants to benefit from the same improved policy literacy, understanding of trade-offs and possible improvements in social cohesion, although validation of such approaches at scale in practical implementations would be essential \cite{van_der_does_small-scale_2023}.

Offline implementations would face real-world barriers to scale this including organizing, coordinating and paying for venues, security and moderators, even in a non-representative distributed effort. In a digital context these specific challenges are replaced with server capacity and possibly a more scalable AI-assisted moderation model, which are much more affordable.
Although ensuring attendance would remain challenging, online systems could eliminate geographic barriers and allow a better mixing across states.

This reduced overhead would allow deliberative experiments at scale, not only at a national level, but also at a more modest level, but still with much lower risks than would be the case with an in-person setting.

\subsubsection{Analytics}
By organizing deliberations online, platforms can maintain accurate recordings and transcripts. These systems would easily be able to also capture granular interaction data, including attributed conversations, queue dynamics, system- or supervisor-generated nudges (prompts for quiet participants) and their outcomes, which already enable valuable research.

The more granular and real-time access to outcome development already has enabled the development of methods to extract ``Superquestions'' that represent clusters of questions across deliberating small groups, that can be posed minutes after finishing to the plenary expert panel \cite{de_question_2025}.

\subsubsection{Interventions}
\label{sec:interventions}
Online deliberative platforms expand the scope of the kinds of interventions a moderator, human or not, has available. Depending on the moderation principles that the developer aims to adhere to, these possibilities should be selectively used.
Unlike offline settings, where interventions are often limited to public (verbal or non-verbal) cues, digital environments have the ability to expose participants to asymmetric interactions tailored to specific individual behavior. For instance, individual participants can be encouraged privately to speak up or contribute, without the need to do this publicly in front of the rest of the group. On our platform we implement this through individual speaking nudges.

Another implementation is allowing participants to engage with the organizers through a helpdesk system, while still listening to and engaging with the discussion.

The integration of real-time analysis complemented with AI could also enable further data-driven interventions. Our platform employs real-time natural language processing to suggest specific arguments in individual speaking nudges that have so far been least discussed.

In our design, we have chosen a more limited set of interventions than theoretically possible, as demanded by our selected neutral and observational moderation strategy. These advantages may be even more meaningful with different moderation strategies.

\subsubsection{Why not online?}
Despite its advantages, online deliberation is not universally the most suitable option available to organizers.
While the approach gains efficiencies by freeing up time spent on travel, coffee breaks, lunches and the night spent between session days -- these moments also offer very tangible benefits such as opportunities for organic social bonding.
It is at these informal interactions during the coffee break and at shared meals, that participants can find camaraderie and further mutual understanding in ways that we have not been able to replicate in virtual environments.
This may for some deliberative settings be a desirable characteristic, worth the added cost and time.

The lower barrier to entry may allow invitees to commit more easily to participating. This comes with the limitation that lower required commitment and ability to track follow-through, may also contribute to volatility in attendance rates.

Furthermore, while digital platforms mitigate a number of constraints for some populations (e.g.\ caregivers, remote workers), they may also introduce, reproduce or exacerbate disparities for others. Digital technology may open doors for many attendees, but it poses a barrier for invitees with limited technological literacy or impairments inadequately addressed by the platform interfaces. While the opportunity to attend from home may help a lot of attendees to satisfy personal constraints with family and work obligations, others may not have the desirable quiet space available in or near their home to effectively participate.

Many of these challenges require a mitigation strategy that could for example include stipends, training and documentation. Stipends could be used to allow participants to buy high-speed internet access, obtain child-care or even rent a space or equipment. Appropriate training and documentation can help attendees to familiarize themselves with a likely new online environment, and get comfortable and be well-prepared to engage in a productive conversation.

\subsection{Previous Work}
\label{sec:literature}
Deliberative mini-publics describe a broad range of methods that use samples of the population to discuss a complex topic, often with the purpose to arrive at recommendations or an understanding how the broader public that they represent, feels about that topic. We kindly refer to existing publications such as by \citet{ryan_defining_2014} for a broader description of the various methods that can be included under the umbrella of deliberative mini-publics, and how they can be categorized.

Deliberative Polling was introduced in the 1990s by James Fishkin and was proposed as a more advanced method to let us know what the hypothetical poll of informed opinions of a population would tell \cite{fishkin_voice_1997, fishkin_deliberative_2000}. This was subsequently implemented in a range of in-person settings, where a representative sample of the population would be brought together to discuss an issue under what Fishkin later defined as \textit{good conditions}.

Such a set of good conditions should allow this diverse set of participants to discuss arguments from both sides of the proposals, gain knowledge and shift their opinion, without allowing inequities to distort the dialogue. He implemented this by gathering random samples as much as possible, engaging the participants in small-group discussions with moderators that are trained not to express any of their own opinions. Carefully crafted and balanced briefing materials formed a shared foundation for these conversations, but there was also an opportunity to engage with experts on the matter. Participants were not asked to arrive at shared conclusions, but rather were polled both before and after the deliberation, providing insights into their opinion shift and knowledge gain \cite{fishkin_democracy_2019}.

\subsubsection{Online Deliberation}
Some work has discussed this specific online platform from certain angles or analyzed specific aspects. This chapter will focus more on painting the big picture and setting the stage, providing hopefully a solid foundation for more future analysis. In particular, we aim to expand on how the platform has been designed, and how it performs in a practical sense. This may be of particular value to those researchers that want to interpret the data from deliberations that are held on the platform.

Online video conferencing has already existed for about a century, and commercial video conferencing platforms have been available for many decades. Both advances have been celebrated and shortcomings criticized in the relevant literature, as the technology is being used for a wide variety of purposes since at least the eighties \cite{egido_video_1988}.

This platform was not the first attempt to create an online environment where deliberation could be scaled effectively, while being truly deliberative to the standards that would be needed. Much of the work discusses online platforms that have a deliberative element but are not designed for deliberative mini-publics, such as social media platforms. For example, \citet{friess_systematic_2015} describe the debates in the literature whether anonymity or identification of participants is advantageous and conclude that identification would benefit rationality and civility, but may negatively impact participation and inclusion. They identify a few important dimensions for deliberation in online spaces: rationality, interactivity, equality, civility, constructiveness and a reference to the common good. It is unclear to what extent the lessons from these broader purpose platforms can be applied to design choices specifically for deliberative mini-publics in a dedicated environment.

The possible designs of `online deliberation platforms' with a focus on mini-publics have been discussed in the literature for at least 20 years. Initially this would for example be implemented through message boards for asynchronous and Internet Relay Chat for synchronous deliberation, although it should be noted that the terminology seems to have been very inclusive of a wide range of deliberative formats \cite{wright_democracy_2007}.

Some studies look at exactly those deliberative mini-publics and what designs would be effective and desirable; \citet{wright_democracy_2007} discuss how some design decisions, including synchronous vs asynchronous, audio vs text and moderation strategy affected the success of early implementations of deliberation platforms. There are several more recent publications that discuss how artificial intelligence could be used to augment \cite{tessler_can_2026, korre_evaluation_2025, mckinney_integrating_2024}, guide or replace a deliberative mini-public \cite{burton_how_2024, tessler_ai_2024}.

There are some earlier platforms described in the literature that have some overlap with the required functionality (see Section~\ref{sec:design}), but often did not reach the scaling benefits \citep[e.g.][]{iyengar_deliberative_2005, luskin_considered_2004, cavalier_deliberative_2009}. Some other studies describe how they managed to organize online deliberations through more widely available commercial video-conferencing software, in combination with human moderators \citep[e.g.][]{herne_can_2025}.

In this chapter, we will focus on outlining the workings of this platform, without making claims regarding how it compares to others. We will leave a more extensive review of the other platforms and how they compare for future work.

\subsubsection{Previous work with this platform}
In the initial demonstration paper of the platform developers the platform and its principles are laid out \cite{fishkin_deliberative_2019} %
and in an expanded version they demonstrated that a deliberation on the platform is possible that achieves parity with an in-person deliberation on a comparable topic \cite{gelauff_achieving_2023}%
.
They demonstrated that the satisfaction of participants was similar to that of participants in in-person deliberations with a comparable participant scope and topic.
Two studies that are currently under review, and whose technical reports are already available, engaged in a randomized controlled trial, where they carefully compared an online with the offline setting.
\citet{gronlund_eduskunnalle_2024} studied a deliberation in Finland where participants were randomly assigned to offline deliberation with a human moderator, online deliberation through a traditional videochat with a human moderator or to online deliberation on this platform without a human moderator, and find similar opinion shifts in all three treatments.
In the second, \citet{reul_nepalis_2026} studied a deliberation in Nepal where participants were assigned to deliberation on this platform with a somewhat different setup (audio-only and a different outcome phase) and demonstrate that these findings also mostly hold in a setting with more technological challenges.

Typically, opinion shift is established by comparing the difference in opinions by the participants with that of a control group. In a study by
\citet{arnesen_knowledge_2024} %
this was established by comparing two treatment groups that discussed different topics, excluding any possible self-selection effects that are platform-specific.

When deliberations take place in structured online environments, this also facilitates more advanced research that would be hard to execute in offline or even less-structured online settings. For example,
\citet{kim_more_2025} %
was able to carefully use the transcripts and participation data matched with the speakers' surveys to study the relationship between demographics and participation as well as opinion shifts.

The potential positive impact of such advances is further enhanced by the possibilities that recent developments in large language models provide. For example, \citet{gelauff_estimating_2024} %
developed an approach to leverage these transcripts in combination with information about the agenda structure and carefully crafted queries to a large language model to estimate the quality of individual contributions to a deliberative conversation. This provides a (somewhat noisy) estimate that does not rely on human annotators, and that can be produced shortly after concluding the event.
\citet{de_question_2025} %
demonstrated that it is possible to rapidly collect the resulting questions for the expert panel from each of the small groups, and create a representative slate of questions that can be answered within the given time by a panel of experts.

A number of publications discuss specific deliberations or research that uses data from such deliberations that took place on the platform. These may provide valuable insights into what is possible and how participants appreciated the deliberative experience -- but often do not discuss the functioning of the platform. We included as many examples as we could track down in Section \ref{sec:deliberations}.

\section{Deliberations on the platform}
\label{sec:deliberations}
The platform has supported deliberative exercises since 2018, initially iteratively developing and testing the design through small-scale pilots, and scaled to accommodate the first deliberations with 100+ participants in 2020 \cite{gelauff_achieving_2023}. As of December 2025, over 30,000 unique participants have engaged in deliberations on the platform, across a wide range of languages, countries and topics.
Table~\ref{tab:deliberations_overview} summarizes large-scale deliberations (100+ unique participants) up until December 2025. The table describes the context (dates, language and countries), the topic and a related external publication, and some high level statistics.

The external publication column is not intended to give an exhaustive overview of publications about these deliberations.

Appendix~\ref{sec:changes} documents the evolution of the platform features and outputs, enabling readers to cross-reference the availability of functionality with specific deliberations in Table~\ref{tab:deliberations_overview}.

\afterpage{\begingroup
\tiny
\setlength{\tabcolsep}{3pt}%
\renewcommand{\arraystretch}{0.85}%
\begin{longtable}{P{0.05\textwidth} | P{0.05\textwidth} P{0.055\textwidth} P{0.075\textwidth} | P{0.14\textwidth}P{0.18\textwidth}| P{0.034\textwidth} P{0.050\textwidth} P{0.050\textwidth} P{0.030\textwidth} P{0.034\textwidth} P{0.034\textwidth}}
\caption{Overview of deliberations on the Stanford Online Deliberation Platform until December 2025 with at least 100 participants. This includes deliberative polls as well as other deliberative exercises, including educational implementations. Related External Publications are a citation of a public report (R.), other research paper (OP.) or a url to a relevant webpage archive (WA.).
Participants counts everyone assigned to at least one room, which is generally larger than the analyzed sample a published report gives for the same deliberation: a report counts those who completed both the pre- and post-deliberation survey, sometimes weighted to the target population. Duration is the scheduled room time a typical participant attends, Participant Hours the time participants actually spent in rooms summed over the deliberation, and Room Hours the scheduled room durations summed. A dash marks information that was searched for and not found; n/a marks deliberations for which no publication is expected.
}\label{tab:deliberations_overview}\\
ID & Date range & Language & Country & Topic & Related External Publication & Rooms & Sessions & Duration (h) & Part. & P. Hours & Rm. Hours\\
\endfirsthead
\multicolumn{12}{l}{\textit{Table \ref{tab:deliberations_overview} continued from the previous page}}\\
ID & Date range & Language & Country & Topic & Related External Publication & Rooms & Sessions & Duration (h) & Part. & P. Hours & Rm. Hours\\
\endhead
202004A & Apr 2020 & ja & Japan & Solar power in Tokyo toward 2030 & WA. \cite{asian_energy_studies_centre_online_2020} & 15 & 3 & 2.8 & 157 & 256 & 26 \\
202005A & May 2020 & en & United States & University policies, healthcare & R. \cite{senior_shaping_2020} & 20 & 3 & 3.4 & 200 & 441 & 46 \\
202007A & Jul 2020 & en & United States & Healthcare, Economy & R. \cite{luo_young_2020} & 40 & 2 & 3 & 289 & 609 & 84 \\
202007B & Jul 2020 & zh-hk & Hong Kong & Smart energy transitions, Greater Bay Area & OP. \cite{cheung_understanding_2022} & 15 & 3 & 2.8 & 180 & 345 & 31 \\
202101A & Jan 2021 & en & United States & School of Sustainability & R. \cite{center_for_deliberative_democracy_deliberative_2021} & 27 & 2 & 3 & 248 & 552 & 60 \\
202103A & Mar 2021 & es & Chile & Constitution of Chile & R. \cite{fishkin_lxs_2021} & 45 & 5 & 7.5 & 564 & 2709 & 245 \\
202103B & Mar 2021 & en & Canada & Canadian Foreign Policy & R. \cite{fishkin_foreign_2021} & 44 & 4 & 6 & 189 & 399 & 74 \\
202103C & Mar 2021 & zh-hk & Hong Kong & Energy & -- & 112 & 1 & 0.7 & 773 & 470 & 44 \\
202104A & Apr 2021 & en & United States & Gene drives & -- & 30 & 1 & 1.2 & 308 & 267 & 21 \\
202105A & May 2021 & en & United States & Electoral College; climate and economic inequality & R. \cite{beyene_shaping_2021} & 120 & 2 & 3 & 727 & 1222 & 114 \\
202105B & May 2021 & en & United States & School of Sustainability & R. \cite{fishkin_new_2021} & 25 & 3 & 3.6 & 218 & 521 & 56 \\
202106A & Jun 2021 & en & United States & Economic inequality & n/a & 40 & 2 & 2 & 356 & 281 & 34 \\
202109A & Sep 2021 & en & United States & Climate and Energy & R. \cite{noauthor_america_2021} & 132 & 4 & 5 & 1262 & 4219 & 414 \\
202111A & Nov 2021 & en & United States & Social Media, Civic Responsibility & R. \cite{acuna_lacarieri_voices_2021} & 64 & 2 & 2.5 & 294 & 485 & 48 \\
202111B & Nov 2021 & en & United States & Race and division in education & n/a & 32 & 1 & 1 & 146 & 125 & 13 \\
202205A & May 2022 & es & Peru & Citizen participation & WA. \cite{rpp_idea_2022} & 25 & 2 & 3 & 203 & 396 & 37 \\
202206A & \multicolumn{1}{l}{Jun/Nov 2022} & no & Norway & Climate, Various policy issues & \mbox{R. \cite{andersen_interim_2022}} \mbox{OP. \cite{arnesen_knowledge_2024}} & 27 & 4 & 4.3 & 224 & 381 & 39 \\
202207A & Jul 2022 & en & United States & Energy and climate & R. \cite{valentine_testing_2022} & 27 & 1 & 1.5 & 139 & 192 & 24 \\
202207B & Jul 2022 & zh-tw & Taiwan & Climate, energy & OP. \cite{fan_effects_2025} & 25 & 2 & 3.2 & 113 & 237 & 23 \\
202207C & Jul 2022 & en & United States & Social Media & n/a & 35 & 1 & 1.2 & 288 & 268 & 28 \\
202208A & \multicolumn{1}{l}{Aug/Nov 2022} & en & United States & Climate and Energy & -- & 142 & 2 & 2.5 & 771 & 1197 & 150 \\
202211A & Nov 2022 & en & United States & Campus and Social Media, Democratic Reform & R. \cite{udaygiri_virtual_2022} & 42 & 2 & 2 & 282 & 368 & 41 \\
202212A & Dec 2022 & 19 languages & 32 countries & Bullying and Harassment in the Metaverse & R. \cite{fishkin_meta_2023} & 890 & 4 & 5 & 7863 & 19060 & 1736 \\
202302A & Feb 2023 & en & United States & Various policy issues & R. \cite{the_goldman_school_berkeley_public_policy_california_2023} & 90 & 4 & 5 & 997 & 3401 & 331 \\
202303A & Mar 2023 & en & United States & Income gap & -- & 40 & 1 & 1.3 & 139 & 160 & 32 \\
202305A & May 2023 & en & Worldwide & Algorithms, Literacy & -- & 55 & 3 & 3 & 266 & 373 & 37 \\
202305B & May 2023 & en & Worldwide & Misinformation & R. \cite{deliberative_democracy_lab_deliberative_2023} & 12 & 2 & 2.5 & 108 & 141 & 15 \\
202306A & Jun 2023 & en & United States & Democratic Reform & R. \cite{deliberative_democracy_lab_results_2023} & 90 & 5 & 5.5 & 706 & 2327 & 225 \\
202310A & Oct 2023 & en & United States & AI in High Schools & R. \cite{ciesla_national_2024} & 20 & 2 & 2.3 & 113 & 152 & 16 \\
202310B & Oct 2023 & 4 languages & 4 countries & Generative AI & R. \cite{chang_meta_2024} & 166 & 4 & 3.8 & 1602 & 3243 & 328 \\
202311A & Nov 2023 & fi, sv & Finland & Drug policy; fuel and motoring taxation & R. \cite{gronlund_eduskunnalle_2024} & 28 & 5 & 5.8 & 266 & 934 & 80 \\
202311B & Nov 2023 & en & United States & AI in the workplace, Reproductive Access & n/a & 30 & 2 & 2 & 181 & 245 & 35 \\
202403A & Mar 2024 & en & United States & Technology and data in the workplace & R. \cite{schrerer_what_2025} & 30 & 3 & 3.8 & 168 & 252 & 25 \\
202403B & Mar 2024 & en & United States & Electoral College & n/a & 163 & 1 & 1.3 & 385 & 179 & 22 \\
202403C & Mar 2024 & zh-tw & Taiwan & AI regulation & R. \cite{taiwan_ministry_of_digital_affairs_utilizing_2024} & 60 & 2 & 2 & 468 & 582 & 56 \\
202408A & Aug 2024 & 6 languages & 6 countries & Anti-Microbial Resistance & R. \cite{mendelson_understanding_2025} & 170 & 3 & 3.8 & 1495 & 3563 & 301 \\
202409A & Sep 2024 & zh-hk & Hong Kong & Climate Change & R. \cite{cheung_civil_2024} & 28 & 2 & 2 & 221 & 368 & 33 \\
202410A & Oct 2024 & 4 languages & 5 countries & AI Agents & R. \cite{chang_meta_2025} & 133 & 3 & 3 & 911 & 1851 & 178 \\
202410B & Oct 2024 & en & United States & Social media regulation for minors & WA. \cite{close_up_foundation_high_2024} & 15 & 1 & 1.2 & 114 & 99 & 11 \\
202410C & Oct 2024 & en & United States & Immigration & WA. \cite{close_up_foundation_university_2024} & 15 & 1 & 1.2 & 159 & 159 & 8 \\
202412A & Dec 2024 & en & United States & Social media regulation for minors & WA. \cite{close_up_foundation_middle_2024} & 15 & 1 & 1.3 & 159 & 159 & 8 \\
202502A & Feb 2025 & ar & Tunisia & Waste Management & -- & 25 & 4 & 5 & 263 & 644 & 58 \\
202504A & Apr 2025 & en & Australia & Nuclear Power & -- & 15 & 1 & 1 & 101 & 100 & 12 \\
202505A & May 2025 & ne & Nepal & Agricultural policy & R. \cite{reul_nepalis_2026} & 40 & 1 & 1.5 & 268 & 315 & 28 \\
202507A & Jul 2025 & en & United States & AI for teenagers & -- & 51 & 3 & 3 & 410 & 937 & 95 \\
202507B & Jul 2025 & en & -- & AI and health & WA. \cite{pmnch_algorithm_2025} & 20 & 3 & 3.8 & 299 & 295 & 24 \\
202508A & Aug 2025 & ja & Japan & Plastic use & -- & 20 & 1 & 1 & 147 & 118 & 13 \\
202509A & Sep 2025 & en & United States & AI policies & R. \cite{deliberative_democracy_lab_national_2025} & 30 & 2 & 2.5 & 186 & 364 & 40 \\
202510A & Oct 2025 & en & United States & Brain-Computer interface & -- & 22 & 3 & 3.8 & 125 & 198 & 29 \\
202510B & Oct 2025 & en & United States & Civil Rights and Responsibilities & n/a & 682 & 1 & 1.2 & 3028 & 2486 & 406 \\
202511A & Nov 2025 & en, hi & India, United States & AI Agents & R. \cite{deliberative_democracy_lab_industry-wide_2026} & 68 & 2 & 2.5 & 535 & 903 & 88 \\
202511B & Nov 2025 & en & United States & Economy; free speech; immigration; affirmative action & n/a & 10 & 4 & 5 & 204 & 354 & 38 \\
202511C & Nov 2025 & en, hi & India, United States & Frontier AI; AI wearables & R. \cite{deliberative_democracy_lab_releases_2026} & 64 & 4 & 5 & 600 & 1018 & 95 \\
202512A & Dec 2025 & pt & Brazil & AI governance and procurement in public legal aid & -- & 50 & 2 & 2.7 & 226 & 396 & 35 \\
\end{longtable}
\endgroup
}

\subsection{Definitions}
\label{sec:definitions}
In this section, we briefly define the terminology that may be used in the tables, exports and data files that the platform produces. This is a list of some definitions that might be more broadly useful:
\begin{itemize}
    \item Room: The smallest deliberative unit. Typically contains 8-15 participants, and discusses a single agenda for an hour.
    \item Roomgroup: A set of identically configured rooms discussing the same agenda in the same language at the same scheduled timeslot.
    \item Session: One or more roomgroups discussing the same agenda within a Deliberation. Can be in multiple configurations (e.g.\ multiple languages, multiple timeslots and/or multiple privacy conditions)
    \item Deliberation or Event: One or more sessions. The broadest deliberative unit.
    \item Agenda: The structured content that a room discusses. An agenda is linked to a specific language and can be assigned to a roomgroup. For agenda subunits, see also Section~\ref{sec:flow}.
    \item Agenda phase: Section within an agenda. Typically this includes Introduction, Discussion, and Outcome Development phases.
    \item Agenda item: A unit within the discussion phase. An agenda item discusses a well-defined topic, such as a proposition, that has arguments associated with it. A discussion phase typically consists of 3-6 agenda items.
    \item User: Users of the platform include participants, admins, observers and helpers. Uniquely identified by its username. One person may use on occasion multiple usernames, for example when they switch devices or mistype their email address when logging in as a guest.
    \item Participant: a user that is expected to enter a room and contribute to the conversation. Includes Normal Users and Guests.
    \item Normal user: A participant that is logged in with a username and password.
    \item Guest: A participant that is logged in with username, screen name and `real name'.
    \item Admin: A user with additional privileges to configure, start and manage a set of roomgroups and agendas. Can technically contribute to a room, but are not expected to.
    \item Observer: An administrative user that can observe a set of roomgroups, but not contribute.
    \item Speaker: The participant that is currently able to speak in their room. In a typical deliberation there is only one speaker at a time.
    \item Turn: A single contribution by a speaker. This has a starttime and endtime, and typically is associated with a recording and transcript. A participant may have multiple successive turns, if no other participants request to speak. A turn only counts when the participant does not prematurely leave the queue or platform.
    \item Interruption: A brief turn that is concurrent with a regular turn.
\end{itemize}

These are some definitions of relevant metrics:
\begin{itemize}
    \item Speaking time: The sum of the turns of a participant within a room.
    \item Speaking turns: The number of turns of a participant within a room.
    \item Allocated time: Amount of time that an agenda item is expected to take. In publications, this is the normalized initially allocated time by the organizer in seconds. In the database, this is not normalized, and values for agenda items should be considered in relation to each other. During the deliberation, the time is re-allocated at several moments.
    \item Room duration: the difference between the scheduled starting time and the scheduled end time of a room. The expected `clean deliberation time' is often a bit less.
    \item (Unique) Participants: the number of participants that was assigned to at least one room in the deliberation. %
    \item Duration: The sum of the room durations for each session; the amount of time that the typical participant is expected to spend in a room.
    \item Participant Hours: the sum of the actual time spent by participants in the rooms over the course of a deliberation.
    \item Room Hours: the sum of the room durations in a deliberation
\end{itemize}

\section{Design of the platform}
\label{sec:design}
Traditional Deliberative Polls require a representative sample (typically 200-300 people) for an in-person event. They would typically over the course of multiple days discuss the materials together, alternating between small-group discussions and plenary expert sessions.
Facilitators for these small-group discussions would be instructed to maintain a neutral, procedure-focused moderation style, avoiding content rephrasing, interim polling or other consensus-seeking interventions.

As a starting point, a few constraints were selected based on realistically available implementation scenarios:

\begin{itemize}
    \item Synchronous interaction: encourage real-time dialogue dynamics
    \item Audiovisual primacy: prioritize video/audio channels to approximate face-to-face interactions, and ensure a single channel of communication.
    \item Small groups: optimize the platform for small discussion groups of 5-20 participants, allowing for both intimacy and diversity of opinions.
    \item Modular agenda design: support many different agendas that can be reused.
    \item Parallel sessions: create a scalable infrastructure for concurrent discussions across cohorts.
\end{itemize}

Technical implementation priorities shaped additional architecture choices. Browser-based access was selected over application-based access to ensure cross-platform compatibility and minimize device dependencies. A minimalist interface design prioritized core deliberative functions over feature bloat to reduce cognitive load for users. Configurability allows the platform to support a wide range of deliberative designs by allowing the organizer to configure their implementation, including the discussion language.

Our platform's design takes inspiration from James Fishkin's core principle: allow participants to think about the content of the deliberation under good conditions. Most of our moderation design choices follow a few core principles:

\begin{itemize}
    \item Pre-structured conversation allows vetting of arguments and providing a common ground for participants. This may not always be recognized as moderation, but when designed properly, it minimizes the necessary interventions.
    \item Procedural transparency ensures that participants perceive the system as fair and unbiased
    \item Participant-driven agenda control with guardrails provides a balance between allowing participants to take ownership of the conversation and focus on the items of most interest while streamlining parallel conversations.
    \item Equitable participation is encouraged as a matter of fundamental fairness.
\end{itemize}

On top of these moderation principles, the system is designed to keep a human in the loop for interventions. This can be either the group of participants, taking process polls when necessary, or a supervisor from the organizing team that considers escalated issues from reporting systems.

An example of a one-page summary of moderator guidelines that was previously made available by \citet{center_for_deliberative_democracy_deliberative_2026} has been summarized in Table~\ref{tab:guidelines}. These are for example implemented through automated timekeeping and transparent speaker assignment and neutral prompt systems that do not privilege specific viewpoints, while encouraging participants to explore under-discussed pre-vetted arguments.
We complemented those recommendations with how those guidelines are mirrored by the operationalization of the principles within the online platform.

\begin{table}
    \centering
    \small
    \begin{tabular}{P{0.35\linewidth}|P{0.55\linewidth}}
    In-person moderator guideline & Online platform implementation \\ [0.5 ex]
    \hline
        Begin with brief introductions & Structured onboarding sequence with technical instructions and ice-breaker introductions \\
        Do not express opinions & Neutral design avoiding normative language, summaries or position-taking \\
        Do not take hand polls & Polls are anonymous and limited to process decisions \\
        Do not add outside information & The agenda displays pre-vetted arguments only, and when they are relevant to the conversation \\
        Do not correct false statements & The system does not take a position or attempt fact-checking, but does display pre-vetted information at the relevant time to all participants. \\
        No need for consensus & The interface and instructions encourage open-ended exploration, consideration of alternate viewpoints and avoid peer pressure.\\
        Ensure balanced opinions & Balanced agenda design. Smart nudges encourage discussing underrepresented pre-vetted arguments. \\
        Encourage everyone to talk & Monologues are capped, speaking queue provides easy entry and speaking nudges prompt quiet participants \\
        Make sure that participants are talking to each other & No moderator persona; participants naturally talk to each other \\
        Less intervention is better & Minimalist interventions: system creates scaffolding for a conversation and intervenes as little as possible. \\
        Maintain good climate for deliberation & Automated toxicity detection flags harmful speech; participant reporting provides escalation pathway to human supervisors
    \end{tabular}
    \caption{Translation of the in-person moderation guidelines of \citet{center_for_deliberative_democracy_deliberative_2026} into platform design principles.}
    \label{tab:guidelines}
\end{table}

Our first designs were based on optimizing for the exchange of arguments between a group of equal participants, and iteratively added scaffolding and moderation actions when necessary after testing. Early prototypes were tested in small-group deliberative exercises under controlled conditions, surfacing for example the need for a helpdesk-style support system for participants with technical challenges, the need for a single stream of discussion without distracting chat-options, and a preference for a transparent queueing system over a more opaque but slightly more equitable weighted-queueing approach.

Through the structured agenda, the system was able to provide topical guidance to the group by providing relevant audio and video prompts and selectively displaying relevant arguments from that agenda next to visual focus points on the screen. This proved to be sufficient in the majority of cases to focus the conversation on the relevant pre-defined topics on the agenda, without requiring a persona to ensure that the conversation would remain on-topic.

This resulted in a single stream of conversation with one speaker at a time and transparent queueing for other participants. While a structured agenda would help the participants to stay on topic and have a common frame of conversation, the system would not limit the conversation artificially. The progression through the agenda was primarily the responsibility of the group, although early pilots quickly demonstrated the need for guard rails to ensure that the groups would finish on time.

\begin{figure}
    \centering
    \includegraphics[width=0.8\linewidth]{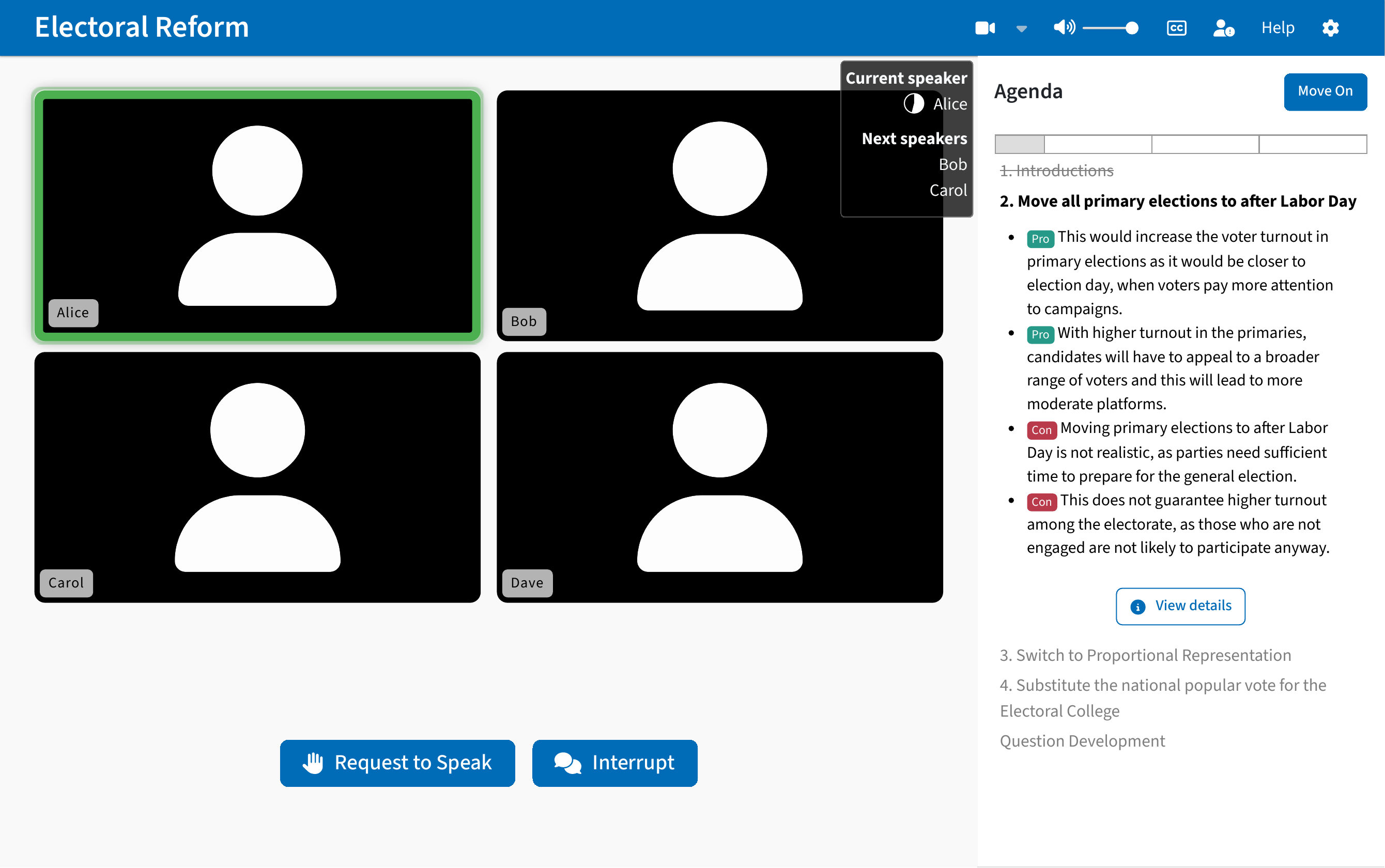}
    \caption{The main user interface of the Stanford Online Deliberation Platform.}
    \label{fig:user-interface}
\end{figure}
The platform's interface integrates the
functionality of the main user interface in three sections: a video panel displaying the current speaker, other participants and the speaking queue; an interaction panel with the various interaction controls for speaking, reporting and note-taking; and the agenda panel visualizing discussion progress and relevant arguments (see Figure~\ref{fig:user-interface}). Early pilots identified 8-15 participants as the desirable group size for optimal participant dynamics, which is what the interface subsequently has been optimized for. Later implementations provided the nuance that this optimum is likely culturally, demographically and topically dependent.

Administrative roles include helpers/observers, who primarily monitor discussions with limited rights and escalate issues via external communication channels, and administrators, who configure rooms, set permissions and access advanced analytics tools.
Critical tools for supervising deliberations are available through a dedicated administrator-interface where they can closely monitor and manage the rooms through the Roomgroup Monitor, a Room Monitor and a Server Monitor, access the Transcript page with transcripts and conversation statistics for each room, aggregate outcomes and export data.

Most of this chapter will focus on the participant-facing interface, but some elements of the administrator-facing interface will be discussed in the context of monitoring, interventions and outcome development.

\subsection{Deliberation Flow}
\label{sec:flow}
The deliberative exchange between participants on the platform happens within the context of a \textit{room}. Participants can be assigned to rooms in advance or on arrival by the system (see Section~\ref{sec:room-assignment}).
Participants experience the deliberation from the moment that they arrive in a room through a few phases, depending on the exact configuration:
\begin{enumerate}
    \item Pre-session: Waiting for other participants to arrive.
    \item Onboarding:
    \begin{enumerate}
        \item Welcome video (optional), allowing the organizer to contextualize the conversation and welcome participants
        \item Instruction sequence, explaining the functionality of the platform
    \end{enumerate}
    \item Ice breaker: participants are invited to introduce themselves to each other and encouraged to try the speaking functionality
    \item Substantive deliberation: each discussion item can be introduced with a short video (optional) and will be preceded by a recorded audio-prompt. Participants then discuss the discussion item while the relevant portion of the agenda is explicitly highlighted on the side.
    \item Outcome Development: A 15-minute structured process to arrive at outcomes for the group (typically questions for a plenary panel). Optional.
    \begin{enumerate}
        \item Submission Phase, where participants can submit a question to be submitted to a plenary session on behalf of their group
        \item Ranking, to prioritize the questions and determine the order in which they are discussed.
        \item Editing, where the other participants can suggest improvements to the question, and the original submitter can edit the question.
        \item Voting, where participants rank the questions twice, to select two questions to be submitted.
    \end{enumerate}
    \item Post-deliberation transition: redirect to external plenary session (optional)
\end{enumerate}

Some time before the start time (typically an hour), the rooms are prepared by the system, after which changes can no longer be made to the setup.

Configuration options include: disable instruction sequence and introductions, set a welcome video, set per-topic videos, disable outcome phase or replace the outcome type (questions) with another type.

\subsection{The use of Artificial Intelligence}
\label{sec:AI}
The rapid advancement of AI-driven tools presents opportunities and fundamental questions for online deliberation platforms. Large language models have already demonstrated the potential to actively intervene in automated ways and steer conversations, or even replace the participants \cite{tessler_ai_2024, korre_evaluation_2025, burton_how_2024}. Our platform embeds AI within the context of the principles that we discussed above, and the overarching objective to enhance human agency in deliberation, rather than replace it.

We anchor this by maintaining deliberation as fundamentally a human-centric process. Participants need to be supported to exchange arguments, views and experiences with other participants.
We prioritize the group's self-governance, allowing them to take ownership of the process rather than to defer to a moderator persona.
Finally, we take a rather guarded approach to implementing AI when it could impact the conversation in an irreversible way. While experiments with analyzing the outcomes of deliberations using AI are relatively safe, we require a human in the loop for most AI-interventions in live discussions.

Within these constraints, we have been using AI tools to make the deliberations more effective and efficient. We use AI to make live transcripts available, and to provide estimates for toxicity and relevance of a contribution, and to keep track which arguments have been most discussed.
Contributions that are estimated to be toxic are flagged to human supervisors, who can then decide whether to intervene and how.
The least-discussed arguments in an agenda item are included in some of the speaking nudges; subtle interventions that help the group to include more perspectives, but effectively reuse pre-vetted materials.

In the preparation and analysis of a deliberation, we see a lot of potential for the use of AI models to scale deliberations even faster. For example, in the preparation, we can use generative AI to more rapidly help prepare draft agendas that human experts can then vet. In the analysis of deliberations, we have been able to use AI to optimize the slate of questions presented to the expert panel, so that the questions best represent the diversity in the submitted questions by the rooms \cite{de_question_2025}.

While the careful implementation of further AI-support for the deliberations will advance, this will be primarily to support, and not replace, the human participants in the process.

\subsection{Room Functionality}
\label{sec:functionality}

\subsubsection{Room Assignment}
\label{sec:room-assignment}
At the core, a participant can participate in a room when they arrive at the designated starting time at the room URL. In the initial design, the platform assumed that the administrator of the deliberation would send the link of the (randomly assigned and/or well-distributed) room to the individual participants. This turned out to be impractical for many reasons, first and foremost because it turned out to be hard to predict accurately how many participants would actually show up -- making it difficult to create rooms with comparable size.

For this reason, the concept of a roomgroup was introduced: a set of identical rooms. When a participant arrives at a roomgroup URL, the system algorithm randomly picks participants from those that have arrived to fill successive rooms (following the alphabet) with randomly picked participants. If the number of unassigned arrived participants is low, the algorithm waits until the number exceeds the minimum size of the room. Once the `ideal size' is achieved, the algorithm moves to the next. This mostly assures a balanced distribution between rooms.

In most cases, a deliberation will consist of multiple sessions. In that case, it turns out to be confusing to participants to send multiple links. We created a `Live Page': an Attendee Dashboard where the participants can see which sessions they are signed up for, when they happen and in what language. A few minutes before the starting time, this is where the roomgroup URL will become available. After a maximum assignment time, the platform will stop assigning participants, and only admins can manually assign very-late arrivals.

If all participants show up on-time, on-arrival assignment would be entirely random. Usually, most participants arrive on time and are assigned together. Frequently, there is a meaningful number of participants that are assigned to rooms after that moment. This is why late-arriving participants can end up being assigned to the same rooms late in the alphabet.

In the Roomgroup Monitor, administrators can monitor the arrival of participants in roomgroups, and their assignment to specific rooms. Administrators have the option to manually assign participants that are still unassigned, which is usually because there are not enough participants to fill a new room to the minimum size, because many more participants arrived than expected, or because they arrived beyond the assignment window. Administrators can also enter an individual room, and manually re-assign a participant if needed (which is by exception needed in cases of unbalanced pre-assignment, or in case of a rare technical or human glitch or an excessive drop-out rate in a specific room).

Configuration options include: time of assignment relative to starting time, minimum room size, ideal room size, time after start time that automatic assignment should end.

\subsubsection{Speaker Management}
\label{sec:speaker-management}
When participants are in a conversation phase of the deliberation (i.e.\ without welcome video, instructions or audio prompt) their microphones are muted and their video screen is miniature sized by default. A participant can click `Request to Speak' and if nobody is currently speaking, they become the current speaker: their microphone is unmuted, their video is enlarged and they get a timer counting down. By default, participants get 45s to speak. When they are done speaking, they can click `Stop Speaking' or after the end of their timer, their turn will end. If someone clicks `Request to Speak' while someone else is speaking, they automatically join a queue that is visible for everyone in the screen. When a speaking turn ends, the first person in the queue becomes the speaker, and the previous speaker can join the queue at the end of the line.

If a participant has an urgent intervention to make, they can click `Interrupt Speaker', which allows them to speak immediately for 15s, at the same time with the speaker. This is used in practice to point out technical (audio) problems or to request an urgent clarification. While they are interrupting, the time of the current speaker is paused. To prevent repeated interruptions, the interrupter is not able to interrupt again for 15s.

Configuration options include: disable speaker queueing, set the speaking time and the interruption speaking time.

\subsubsection{Speaking Nudges}
\label{sec:nudges}
\begin{figure}
    \centering
    \includegraphics[width=0.28\linewidth]{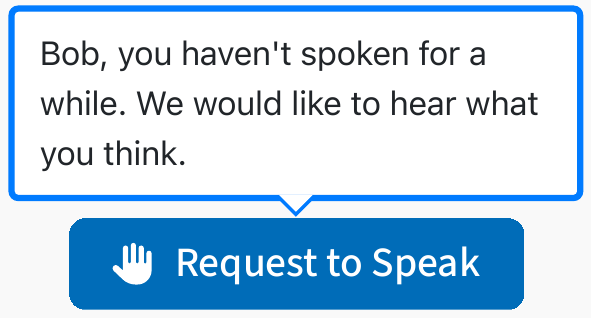}
    \caption{Example of a speaking nudge}
    \label{fig:nudge}
\end{figure}
The system employs some specific nudges to encourage participants to engage with the conversation. Where in an offline conversation, a facilitator can only do this at a group level, in this online setting we can use a more personalized approach. We created a `speaking nudge' that provides a participant with a visual encouragement to speak up, at the same time emphasizing the location of the `Request to Speak' button.

The speaking nudge is activated in 3 situations: 1) when nobody in the room has spoken for 20 seconds, the speaking nudge is sent to everyone at the same time; 2) When a participant has not spoken for a configurable amount of time (typically 180s), that participant is sent a nudge; 3) An admin can manually send a speaking nudge to any individual, provided that they have not already been nudged very recently.

For each language, up to 20 different texts have been prepared from which one is randomly drawn at each nudge occasion. These texts can have placeholders for the participants' screen name and for the least-discussed argument from the current agenda item. Randomly, the system will skip a nudge, which allows us to estimate effectiveness and optimize the nudge texts. Examples in English include: ``What are your thoughts on the proposal being discussed?'', ``\{name\}, you haven't spoken for a while. We would like to hear what you think.'' and ``What do you think about this argument? \{pro/con\}.''

Configuration options include: replace nudge texts with custom texts, time of room silence required to trigger room speaking nudge, time of individual non-speaking required to trigger individual speaking nudge.

\subsubsection{Agenda Management}
\label{sec:agenda-management}
The administrator configures the starting time and duration of the roomgroup, from which its end time follows. The agenda management is based on the assumption that it is important that all rooms end at the designated end time, and that participants should be able to move the conversation forward when they believe the arguments have been discussed. If an outcome phase is selected, a fixed amount of time of 15 minutes is allocated to this. The various audio instructions typically take up another 5 minutes. The administrator assigns a duration to each agenda item, which is understood to include the explainer video and the audio prompt at the start of that agenda item. At the start of each agenda item, the system calculates how much time is available for the remaining agenda items, and after optionally deducting 15 minutes for the outcome development, it distributes that remaining time over the remaining agenda items by ratio of their assigned duration, resulting in an allocated time for the upcoming agenda item.

Participants have access to the `Move On' button, which they can click to initiate a poll asking the room participants privately whether they are ready to move on to the next agenda item. If a sufficient portion of the attendees agrees that this is the case, the system will decide to move on. The threshold for this decision is more than 50\% of the responses, and at least two votes in agreement.

When the group reaches 90\% and 100\% of the allocated time, the system will send the participants a move-on nudge, asking if all arguments already have been discussed, drawing attention to the move-on button. If no consensus has been reached to move on at 120\% of the allocated time, the system will decide itself to move on. An exception is the last agenda item, which will always reach a forced decision to move on at 100\% of the allocated time.

A decision to move on means that the system disables the `Request to Speak' button, lets the current participants in the speaking queue finish, and then the next agenda item will start.

Configuration options include: disable agenda management and the text of the confirmation question to the poll initiator.

\subsubsection{Agenda Display}
\label{sec:agenda-display}
During the deliberation, the agenda is displayed in a dedicated box. The agenda items are collapsed and the current agenda item is expanded to display the relevant pro- and con-arguments. A participant can click on `View details' to see the full table of arguments, click on `previous items' to see the previously completed items and expand any other agenda item to see those arguments (which only affects their personal view). The progression in time is visualized at the top of the agenda as a horizontal bar with sections for each agenda item.

The order of the arguments and the agenda items is typically set by the administrator. The administrator can also select which arguments to display by default in the agenda item, and which are only visible in the full table under `View details'.

Configuration options include: randomize the arguments within-type, replace the pro-con setup with an unspecified setup where the administrator could for example define considerations for a more complex agenda item, that does not have arguments in favor or against.

\subsubsection{Outcome Development}
\label{sec:outcome-development}

\begin{figure}
    \centering

    \begin{subfigure}[t]{0.3\textwidth}
        \centering
        \includegraphics[width=\linewidth]{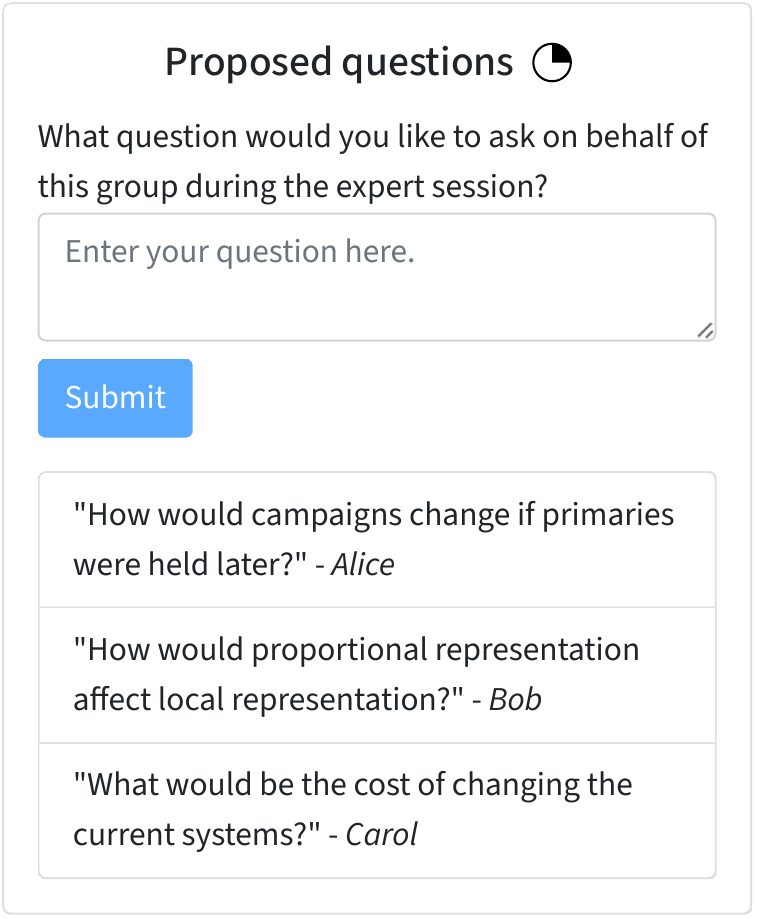}
        \caption{}
        \label{fig:q1}
    \end{subfigure}\hfill
    \begin{subfigure}[t]{0.3\textwidth}
        \centering
        \includegraphics[width=\linewidth]{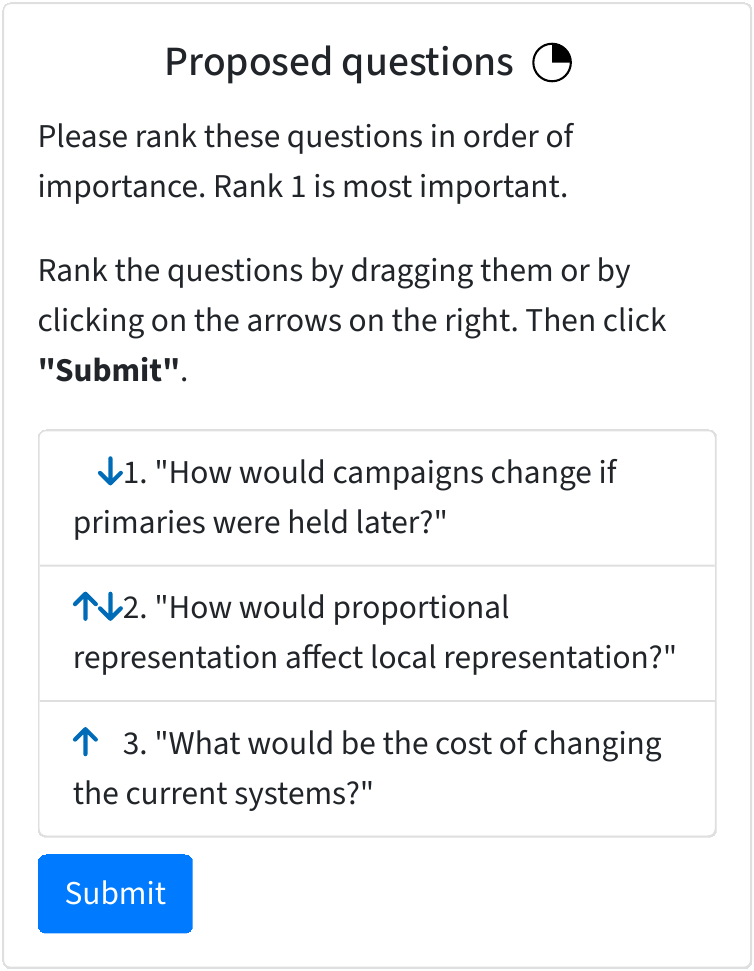}
        \caption{}
        \label{fig:q2}
    \end{subfigure}\hfill
    \begin{subfigure}[t]{0.3\textwidth}
        \centering
        \includegraphics[width=\linewidth]{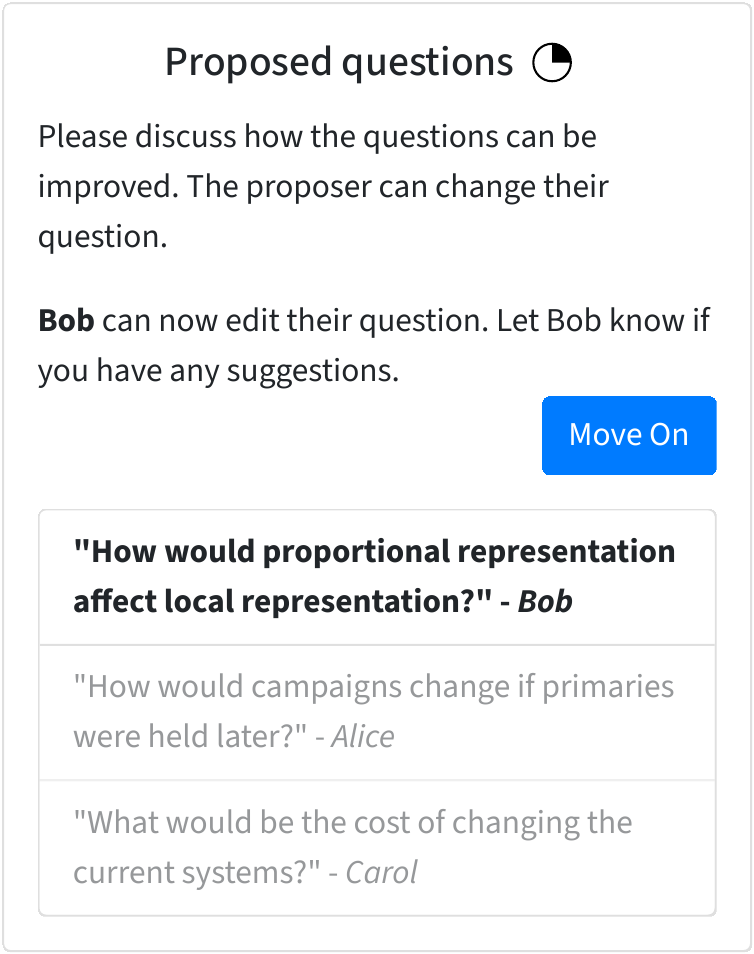}
        \caption{}
        \label{fig:q3}
    \end{subfigure}

    \vspace{0.5em}

    \begin{subfigure}[t]{0.3\textwidth}
        \centering
        \includegraphics[width=\linewidth]{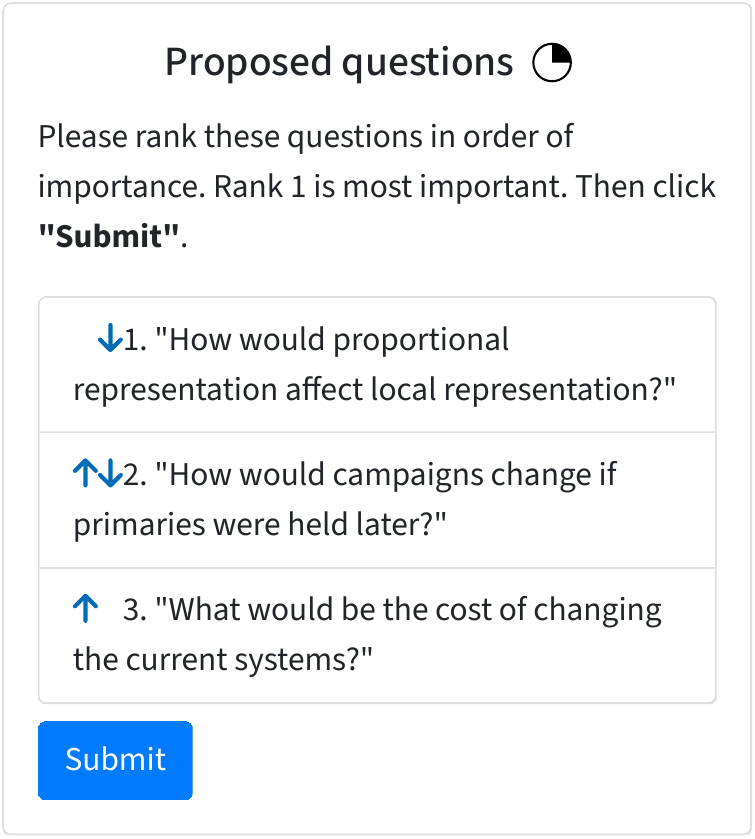}
        \caption{}
        \label{fig:q4}
    \end{subfigure}\hfill
    \begin{subfigure}[t]{0.3\textwidth}
        \centering
        \includegraphics[width=\linewidth]{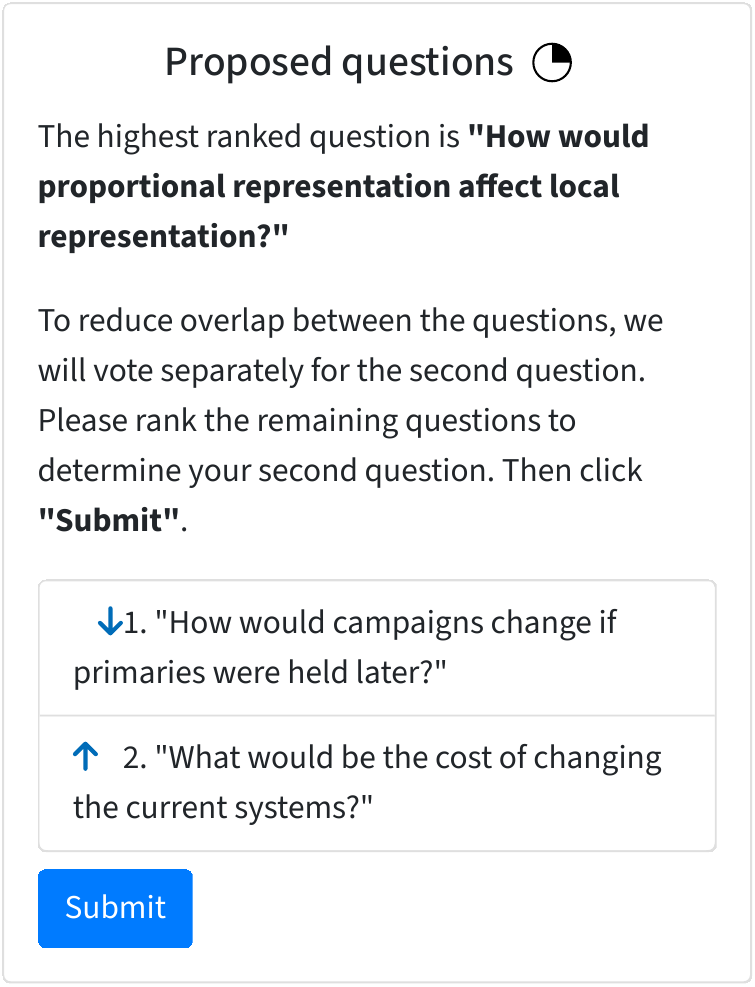}
        \caption{}
        \label{fig:q5}
    \end{subfigure}\hfill
    \begin{subfigure}[t]{0.3\textwidth}
        \centering
        \includegraphics[width=\linewidth]{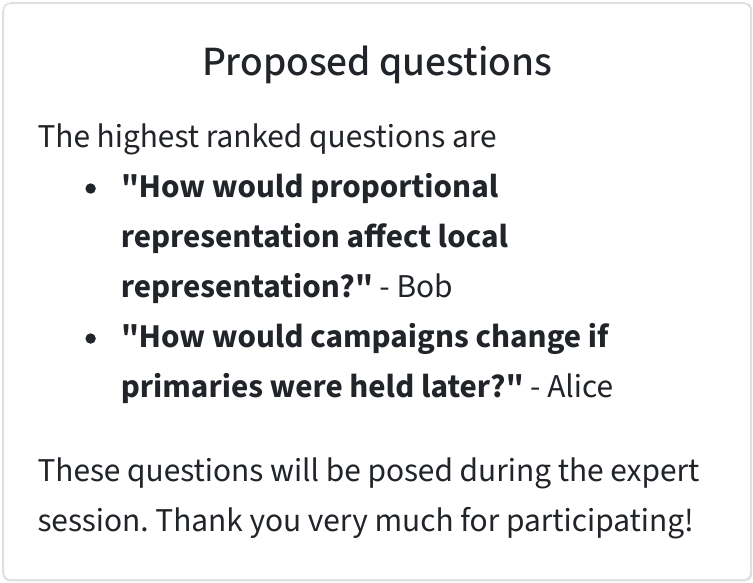}
        \caption{}
        \label{fig:q6}
    \end{subfigure}

    \caption{Outcome development phase. (a) Participants propose questions. (b) Participants rank questions in order of importance. (c) Participants discuss and edit each question. (d) \& (e) Participants rank the questions twice to select two questions to be submitted. (f) The two highest ranked questions are selected.}
    \label{fig:question}
\end{figure}

If an outcome development phase is selected by the administrator, the group will enter it at the end of the last regular agenda item. This is 15 minutes before the scheduled end time, unless the group decides to move on prematurely. The administrator can configure a different outcome type than questions, but for this section, we will assume questions, which is the most common type.

After an introductory explanation by the system, participants are invited to submit one question each in parallel, which will become visible on the side of the screen upon submission (Figure~\ref{fig:question}a). If the question meets a minimum length threshold (language dependent) and is not submitted at the end of the phase, the system will auto-submit it.
After submission is complete, the participants are invited to rank the proposed questions in order of importance, by clicking arrows beside the questions or dragging them (Figure~\ref{fig:question}b). The system will aggregate these rankings using a Borda score \cite{de_borda_memoire_1784} which then determines the order in which they will be edited. The questions appear as agenda items on the side of the screen, and starting with the highest ranked question, the participants are invited to give feedback to the author on the question, and the author has the opportunity to improve the question (Figure~\ref{fig:question}c). When the group determines the question is its best version, they can vote to move on to the next. Because time may run out before all questions are edited, it is important to begin with the highest ranked question. About three minutes before the scheduled end, or when all questions have been discussed, the participants are asked to rank the questions and select which question should be submitted on their behalf to the plenary panel (Figure~\ref{fig:question}d). After the first question is selected, they are asked to do so again for the second question to minimize possible overlap (Figure~\ref{fig:question}e). Aggregation of rankings uses again the Borda score.

\subsection{Real-time Monitoring}
\label{sec:monitoring}
In most deliberations, there will be a few administrators supervising the proceedings, together with a few helpers and typically one or a few helpdesk support members. Administrators generally have the option to intervene in the deliberation, helpdesk support members respond to help queries via the `Help' button that are relayed to Zendesk (a third party helpdesk software) while helpers can only observe the conversations, access the transcripts and report their findings to the administrators or helpdesk support.

The Room Monitor is the central monitoring page for most administrators and helpers, giving them a quick overview of all deliberation rooms, their number of participants, the agenda phase that the room is in, how much time expired since the most recent speaker and warning indicators of any problems that the system may provide as well as access to (invisibly) observe the room conversations live, or through a transcript on the Transcript pages.

\begin{figure}
    \centering
    \includegraphics[width=0.9\linewidth]{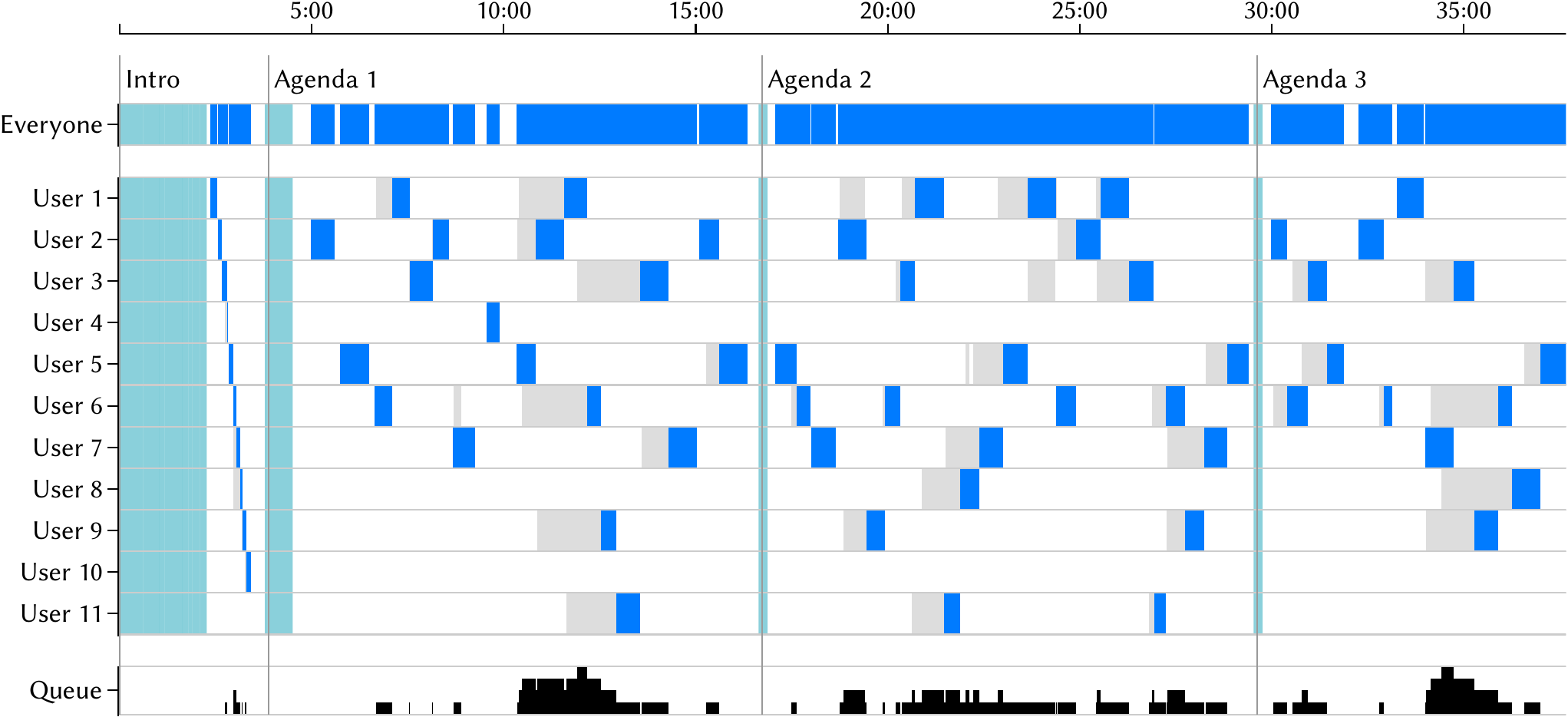}
    \caption{A timeline on the Transcript page visualizes the progress of the conversation.}
    \label{fig:timeline}
\end{figure}

On the room-specific Transcript pages, the administrator and helper can find a log with transcripts and links to audio recordings of the participants' spoken contributions, actions and relevant administrative interventions, up to the moment of loading. The page visualizes the progress of the conversation so far in a graph at the top (Figure~\ref{fig:timeline}), and reports the (intermediate) results of the outcome development at the bottom. The page is also a helpful tool in diagnosing possible technical problems that participants may report by providing some technical information such as when a participant joined, left and rejoined the room, and their browser and device information at the moment of joining as well as audio snippets from their spoken contributions.

\subsubsection{Undesirable Behavior}
\label{sec:undesirable-behavior}
When individual participants are behaving in undesirable ways, this can be picked up through automatic detection, reporting by other participants or via reports from observers and the helpdesk. Automatic detection is always based on automated transcripts that are live generated from the audio streams. The transcript is then used to establish a toxicity score (initially Perspective API; in 2026 this was replaced by a bespoke scoring service). In addition, a simple word filter is defined for each language.

If the word filter is triggered, or if the toxicity score is higher than a threshold, the system will report it as potentially abusive language.
Participants can also report another participant in their room as being abusive, off-topic, having technical issues or report for an `other' reason. If the report is system-flagged, or if the user-report is potentially abusive or off-topic, the system will initiate a poll among the participants that were neither reporter nor reportee. It will then pass the outcome of that poll (or immediately, if the issue is technical or `other'), to the human supervisors as a `flag'.

\subsubsection{Admin Interventions}
\label{sec:admin-interventions}
At the start of a conversation, the administrator has discretion on when to start the room, based on observed quorum. They can do this from the room itself, or from the Room Monitor. During the conversation, the administrator and helper can enter the room without being visible to regular participants. The helper can only observe the conversation (and some additional data), while the administrator has the ability to intervene.

In order to deal with individual problems, they can message a participant (which will pop up as a message window in the browser), send them a speaking nudge, force their browser to refresh or redirect them to a different URL. They can also temporarily prohibit a participant from speaking, or permanently block their access to the platform.

In order to deal with room-wide issues, the administrator can trigger a move-on nudge, initiate a move-on poll and even manually force the outcome of such a poll. The administrator has also the possibility to speak up as a regular participant, at which moment they do become visible to other attendees. In more drastic situations, the administrator can manually `move' the room to a different phase in the agenda (including the very beginning before the room started, allowing an effective `reset') and type a message on the screen.

\subsubsection{Superquestion Generation}
\label{sec:superquestions}
After the rooms finished deliberating, the administrator can aggregate the submitted outcomes using the Superquestion Generator. The algorithm for this generator has been discussed more in-depth by \citet{de_question_2025}, and effectively collects all submitted questions (or the top-$k$ per room), and finds a slate of questions that best represents the clusters of similar questions that were submitted. This could (based on the settings selected) be exact submitted questions, or AI-generated questions that represent clusters of questions.

This allows the administrator to collect and aggregate the submitted questions to a slate that can then be presented to a panel of experts within minutes, which provides a logistical advantage that participants can immediately be redirected to the plenary session, without risking drop-off in between.

\section{Discussion}
The emergence of scalable online deliberation platforms like this platform does not merely constitute a digitization of existing methods -- it redefines the scope of what is possible through democratic innovation. As we outlined in Section~\ref{sec:why-online}, these tools lower barriers to organizing deliberative mini-publics while enabling novel designs for democratic engagement. Yet, more transformative potential lies just around the corner: scalable structured dialogue beyond deliberation, deliberation in education and longitudinal civic analysis.

The modular architecture of the platform already supports several structured conversation formats that go beyond traditional deliberative mini-publics. For instance, the \textbf{use of deliberative discourse as an educational tool} is not a new one. U.S.\ high schools and universities have already demonstrated some of the potential that adopting deliberative methods with students could have in terms of depolarization \cite{siu_plea_2025, khan_how_2023}, and the platform has in fact already hosted several deliberations that were primarily educational in nature. These deliberations do not have to follow the same rigor in terms of representativeness, although for very different reasons diversity becomes important. This is where online platforms may provide a very practical solution, by matching multiple classes at large distances, and ensuring a minimum of diversity in each of the rooms. The minimum required number of participants is more a practical limitation than to satisfy statistical significance.

These educational exercises, as well as many other models such as civics-minded democratic engagement initiatives, prioritize not the report but rather the epistemic empathy and depolarization that students might experience. One of the most encouraging components might be that the participants with a deep interest in politics are not encouraged to focus on traditional debate with participants trying to persuade each other or at least the audience, but rather focus on understanding the other's perspective.

The \textbf{reproducible workflows} that platforms like the Stanford Online Deliberation Platform bring, could enable a systematic comparison across time and populations. Digital iterations allow organizers to repeat deliberations on the same topic, with varied participant cohorts. This reproducibility could transform how we study opinion formation, and make it possible to isolate the impact of variables such as agenda structures, expert input, reading materials and facilitation protocols.

A logical further challenge would be to attempt to disentangle the \textbf{causal drivers of opinion shifts}. While traditional Deliberative Polling with the pre- and post-deliberation surveys already captures attitude changes, this is usually analyzed in the aggregate. When the individual patterns are considered, these opinion shifts become more complex and may require a deeper understanding of the micro-interactions through studying the specific argument exchanges, capturing nonverbal cues and tracking procedural nudges. The interaction logs are one element in the list of requirements that could make this kind of study feasible.

\section{Author Contributions}
Lodewijk Gelauff served as the lead author, responsible for the conceptualization, original drafting, and manuscript synthesis, as well as the product management and strategic design of the platform.
Sukolsak Sakshuwong was responsible for the software architecture and core technical implementation during the platform's foundational development phase; he also provided technical illustrations for this chapter and performed critical revisions.
Ashish Goel provided senior project oversight, secured funding for the infrastructure and development.

\section{Acknowledgements}
This material is partially based upon work supported by the National Science Foundation under NSF Awards \#2333849 and \#1637418.

The authors would like to thank Dr.\ Alice Siu for her vital role in facilitating many of the collaborations with academic and non-academic partners that this platform hosted. She provided and translated key insights into the platform's requirements and opportunities and we are fondly indebted to her role as ``Voice of the Platform'' in English and Chinese. We extend special gratitude to Professor James Fishkin for his foundational work into the social dynamics of deliberative polling and Professor Kamesh Munagala for early-stage conceptual input and grant support.

Finally, we want to thank and acknowledge the essential contributions of the student developers and researchers that were part of the Crowdsourced Democracy Team, that turned visions into code: Harshvardhan Agarwal,
Laura Castro Venegas, Lovish Chopra, Shoaib Mohammed, Bhargav Dindukurthi, Nikhil Garg, Xinyu Hu, Daniel Kharitonov, Isa Liang, Ruochen (Chloe) Liu, Arafat Mohammed, Kamesh Munagala, Liubov Nikolenko, Shouzhuo Sun, Zixin (Coco) Xu and Sravya Yandamuri.

\subsection{AI Disclosure}
AI-generated feedback was selectively applied to improve readability, with all revisions validated by the authors. Final content and analysis remain the authors' own.

\putbib
\end{bibunit}

\newpage
\appendix
\section{Some relevant software changes over time}
\label{sec:changes}
This overview is not intended to be exhaustive, but rather lists changes that the authors think are possibly relevant to organizers and evaluators when interpreting the deliberation data or behavior in comparison with the description in this chapter. It mostly focuses on new functionality, or changes that affect the user interface meaningfully for the participants. The overview may skip features that are not discussed at all in this chapter, bug fixes (unless they changed what participants experienced) and performance improvements. The list covers changes up to January 2026, and later changes only where this chapter discusses them.

\begin{itemize}
    \item 2019-02 Introduce toxicity scoring
    \item 2019-06 Introduce an observer interface
    \item 2019-07 Track agenda interactions
    \item 2019-07 Show the agenda during the question proposal phase
    \item 2019-07 Order participant videos by the most recent speaking turn
    \item 2019-07 Introduce `Free-for-all' mode, in which agenda management and speaker management are disabled
    \item 2019-08 Let the speaking queue finish before the room moves on
    \item 2020-01 Introduce note-taking for participants
    \item 2020-04 Add helpdesk functionality
    \item 2020-04 Add optional subtitles for participants
    \item 2020-04 Introduce symbolic avatars for participants without an active video stream
    \item 2020-04 Automatically move the room on at 120\% of an agenda item's allotted time (100\% for the last item), and reallocate the remaining time across questions during question editing
    \item 2020-07 Randomize the question order per participant during the question ranking phase, and log the initial order
    \item 2020-10 Add an option to play a YouTube video at the start of an agenda item
    \item 2021-01 Include `technical problems' as a reason to report another participant
    \item 2021-01 Make the outcome phase configurable, allowing for outcome types other than questions
    \item 2021-02 Make the speaking nudge more visible by adding a blue border
    \item 2021-02 Introduce roomgroups: groups of identically configured rooms with an alphabetical suffix
    \item 2021-03 Select each nudge text at random from up to 20 texts, rather than always using the same one
    \item 2021-03 Introduce automated random room assignment on arrival
    \item 2021-11 Log partially written questions from the question development phase
    \item 2022-01 Prevent a participant from proposing to move on again immediately after their proposal failed
    \item 2022-02 Improve tracking of room configuration options
    \item 2022-03 Allow organizers to upload improved transcripts
    \item 2022-04 Introduce a non-nudging benchmark by randomly skipping some speaking nudges (partial until 2022-10)
    \item 2022-05 Introduce nudges that refer to specific under-discussed arguments of the current agenda item
    \item 2023-01 Restructure the voting mechanism for reported participants
    \item 2023-03 Add optional pre-discussion questions
    \item 2023-10 Introduce optional room assignment based on the answer to a pre-discussion question
    \item 2023-10 Introduce automated escalation when the toxicity score is high
    \item 2023-11 Split the universal `free-for-all' mode into separate options for disabling speaker management and agenda management
    \item 2023-11 Automatically submit draft questions that meet the minimum length requirement at the end of the question development phase
    \item 2024-01 Allow participants to change their screen name
    \item 2024-03--2025-11 Room assignment did not appropriately consider the pre-discussion answer when the organizer selected this option (rarely used)
    \item 2024-10 Introduce an optional experiment mode requiring larger majorities to move on early in an agenda item
    \item 2024-11 Introduce an experiment mode with alternative wording for the move-on poll
    \item 2025-02--2026-02 Speaking nudges were unintentionally not shown to participants, although they were logged as shown
    \item 2025-03 Migrate video and audio processing from Twilio to Amazon Chime
    \item 2025-04 Upgrade speech recognition to a newer model
    \item 2025-04 Introduce AI-supported generation of a question slate from the submitted questions (superquestions)
    \item 2025-05 Introduce tree-based avatars, rather than symbolic representations, for participants without an active video stream
    \item 2025-05 Introduce optional post-discussion questions for evaluation purposes
    \item 2025-05 Allow the order of pro and con arguments within an agenda item to be randomized (not uniformly random until 2026-01)
    \item 2025-05 Introduce a re-transcription pipeline to improve transcripts after the event
    \item 2025-05 Restructure the audio recordings and transcripts
    \item 2025-05 Log the outcome of each move-on poll, including the threshold and the elapsed time
    \item 2025-07 Introduce an experiment mode allowing questions to be edited in parallel
    \item 2025-10 Skip the ranking and voting phases when too few questions were submitted
    \item 2025-10 Play a sound when an interruption starts and ends
    \item 2025-11 Log when participants turn their camera on or off
    \item 2026-01 Log each participant's room assignment, including the assignment mechanism and the decision made
    \item 2026-02 Adjust the minimum length for automatically submitting a typed question for some languages
    \item 2026-05 Replace the Perspective API with a self-hosted toxicity scoring service
\end{itemize}

\end{document}